\documentclass{PoS}

\title{O(4) scaling analysis in two-flavor QCD at finite temperature and density with improved Wilson quarks}

\ShortTitle{O(4) scaling analysis in two-flavor QCD at finite temperature and density ...}

\author{\speaker{T.~Umeda}$^1$, S.~Ejiri$^2$, R.~Iwami$^3$, K.~Kanaya$^{4,5}$, H.~Ohno$^6$, A.~Uji$^3$, N.~Wakabayashi$^3$, and~S.~Yoshida$^{7}$\footnote{Present address: Los Alamos National Laboratory, USA} \hspace{1mm} (WHOT-QCD Collaboration)  \\ \\
$^1$Graduate School of Education, Hiroshima University, Hiroshima 739-8524, Japan \\
$^2$Department of Physics,~Niigata University, Niigata 950-2181, Japan\\
$^3$Graduate School of Science and Technology,~Niigata~University,~Niigata~950-2181,~Japan\\
$^4$Center for Integrated Research in Fundamental Science and Technology (CiRfSE), University of Tsukuba, Tsukuba 305-8571, Japan\\
$^5$Faculty of Pure and Applied Sciences, University of Tsukuba, Tsukuba 305-8571, Japan\\
$^6$Center for Computational Sciences (CCS), University of Tsukuba, Tsukuba 305-8577, Japan\\
$^7$Key Laboratory of Quark and Lepton Physics (MOE) and Institute of Particle Physics,
Central China Normal University, Wuhan 430079, China \\
}

\abstract{
We study the curvature of the chiral transition/crossover line between the low-temperature hadronic phase and the high-temperature quark-gluon-plasma phase at low densities, performing simulations of two-flavor QCD with improved Wilson quarks.
After confirming that the chiral order parameter defined by a Ward-Takahashi identity is consistent with the scaling of the O(4) universality class at zero chemical potential, we extend the scaling analysis to finite chemical potential  
to determine the curvature of the chiral transition/crossover line at low densities
assuming the O(4) universality.
To convert the curvature in lattice units to that of the $T_c(\mu_B)$ in physical units, 
we evaluate the lattice scale applying a gradient flow method. 
We find $\kappa=0.0006(1)$ in the chiral limit, which is much smaller than that obtained in (2+1)-flavor QCD with improved staggered quarks.
}

\FullConference{34th annual International Symposium on Lattice Field Theory\\
		24-30 July 2016\\
		University of Southampton, UK}

\begin{document}

\section{Introduction}
\label{sec:intro}

Precise location of phase boundaries between the hadronic and quark-gluon-plasma phases 
is important in phenomenological analyses of quark matter produced in relativistic heavy-ion collisions.
In this study, we concentrate on the shape of the chiral transition/crossover line at low densities. 

\begin{figure}[tb]
\begin{minipage}{7cm}
\centering
\vskip -0.3cm
\includegraphics[width=5.8cm]{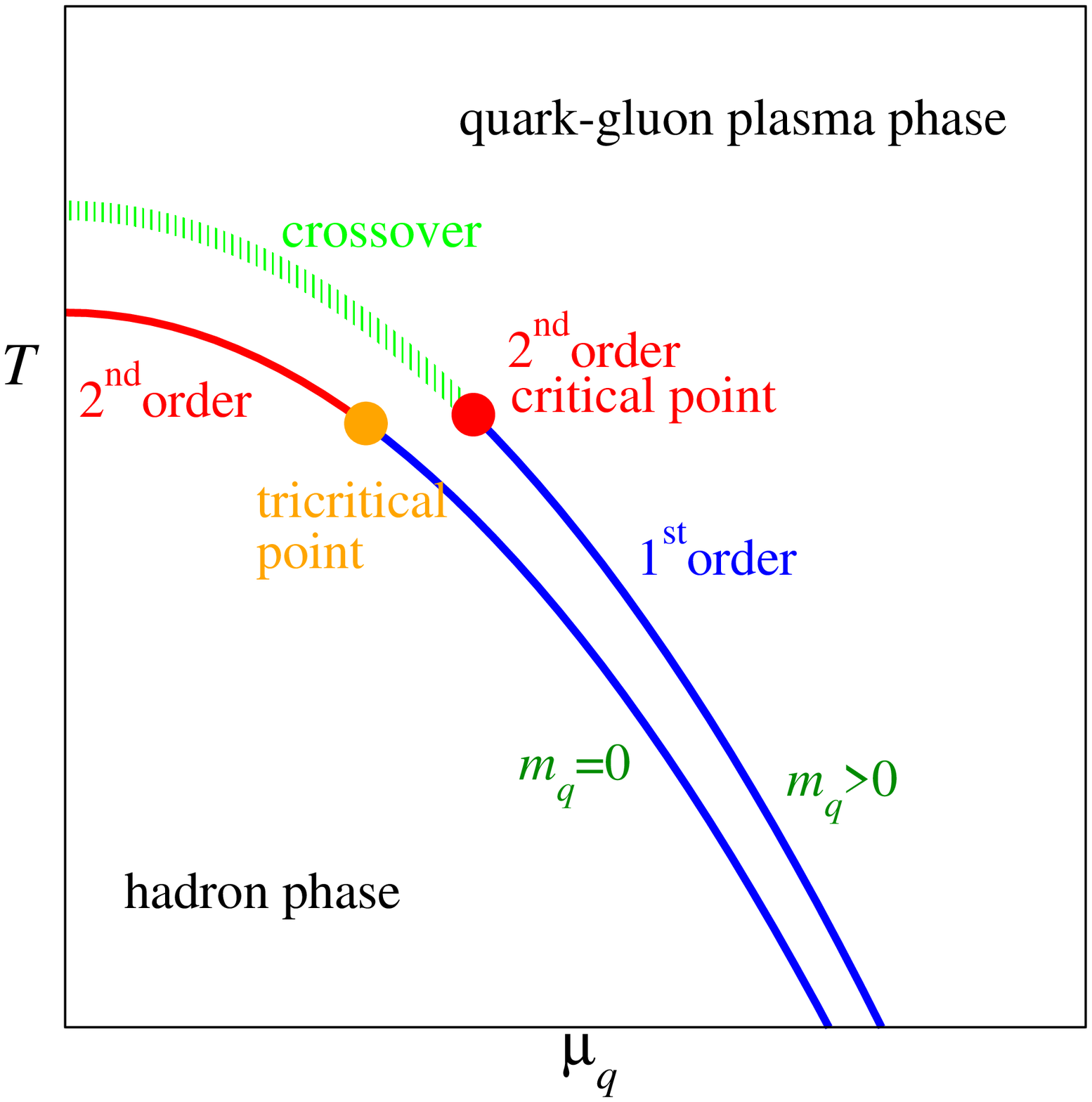}
\caption{
Expected phase daigram of two-flavor QCD at finite temperature and density in the standard scenario.
}
\label{fig1}
\end{minipage}
\hspace{8mm}
\begin{minipage}{7cm}
\centering
\vskip -0.3cm
\includegraphics[clip,width=6cm]{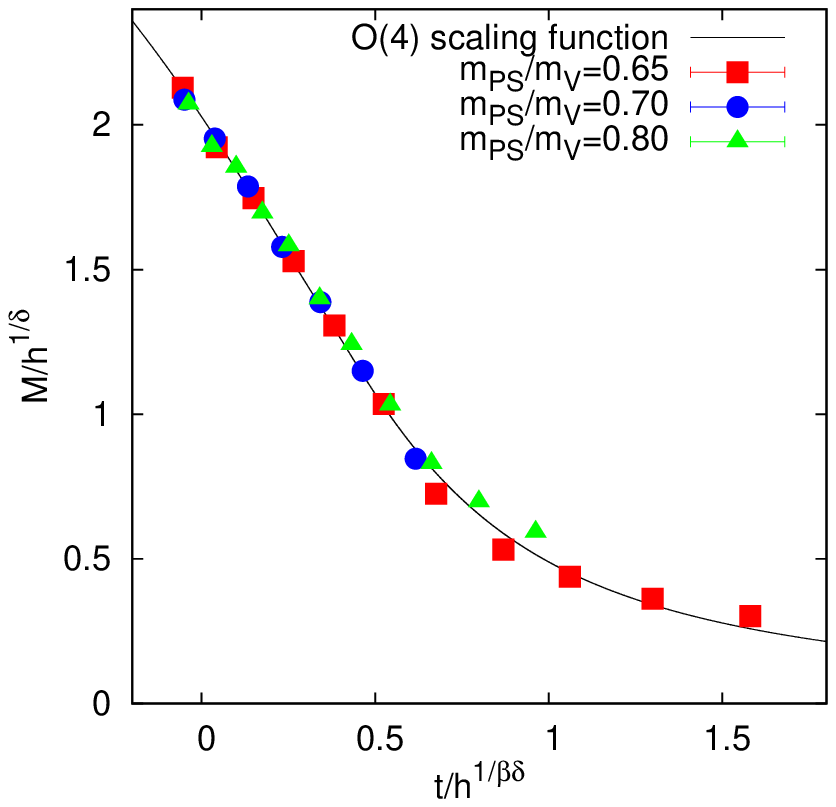}
\vskip -0.2cm
\caption{
O(4) scaling plot of the chiral order parameter. 
}
\label{fig2}
\end{minipage}
\end{figure}

The standard scenario for the QCD chiral transition/crossover, based on the fact that the U$_\textrm{A}$(1) symmetry is explicitly violated by the quantum anomaly at any temperature, predicts that the chiral transition in two-flavor QCD is of second order in the chiral limit $m_q=0$ at small chemical potential $\mu_q$,
and the scaling property around the critical point is universal to that of three-dimensional O(4) spin models. 
When $m_q \neq 0$, the second order transition will turn into a crossover but is expected to change to a first order transition at sufficiently large $\mu_q$.
We illustrate the conjecture in Fig.~\ref{fig1}.
The O(4) scaling behavior in QCD has been reported in Ref.~\cite{iwasaki,cppacs1} for Wilson-type quark actions and Ref.~\cite{bnl-bie09} for an improved staggered quark action.
On the other hand, it was recently argued that 
the chiral susceptibilities of $\pi$ and $\eta$ mesons become the same in the high temperature phase, suggesting an \textit{effective} restoration of the U$_\textrm{A}$(1) symmetry at the chiral transition temperature for these quantities \cite{aoki12}.
If this is the case, the chiral transition of two-flavor QCD may be of first order and the scaling around the critical point at finite $m_q$ may belong to the Z$_2$ universality class of Ising spin systems\footnote{See Refs.~\cite{Perissetto13,Nakayama15} for other possibilities.}.
However, the relation with the explicit violation of the U$_\textrm{A}$(1) symmetry remains unclear and the nature of the chiral transition in two-flavor QCD is still an open problem. 
In this study, we first reconfirm that the QCD data is consistent with the O(4) scaling, and then evaluate the curvature of the chiral crossover/transition line at small $\mu_q$ assuming the O(4) universality.
We reserve a test of the other possibility for the next step.
To avoid deformation of the flavor symmetry and theoretical uncertainties about the continuum limit, we adopt improved Wilson quarks.

\section{Method}
\label{sec:method}

The order parameter of the O(4) spin model is given by the magnetization $M$.
In the vicinity of the second order transition point, $M$ follows the scaling relation
\begin{eqnarray}
M/h^{1/\delta} = f(t/h^{1/\beta \delta})
\end{eqnarray}
with the critical exponents $1/(\beta \delta) = 0.546$ and $1/ \delta = 0.2073(4)$ \cite{engels00,engels03}, 
where $h$ is the external magnetic field, $t=(T-T_c|_{h=0})/T_c|_{h=0}$ is the reduced temperature, 
and $f(x)$ is the scaling function.
In two-flavor QCD at $\mu_q=0$, we may identify $M = \langle \bar{\psi} \psi \rangle$, $h=2m_q a$, and $t= \beta - \beta_{ct}$,
where $\beta_{ct}$ is the critical point of $\beta=6/g^2$ in the chiral limit
and $a$ is the lattice spacing\footnote{Following a convention, $\beta$ denotes both the lattice gauge coupling and a critical exponent.}.
At $\mu_q\ne0$, because the QCD action has the chiral symmetry at $m_q=0$, we expect the same critical properties. 
In the low density region, because of the symmetry under $\mu_q \rightarrow -\mu_q$, the leading contribution from small $\mu_q$ may be incorporated by just replacing $t$ by \cite{bnl-bie10}
\begin{eqnarray}
t= \beta - \beta_{ct} + \frac{c}{2} \left(\frac{\mu_q}{T} \right)^2 .
\label{eq:sclv}
\end{eqnarray}
The coefficient $c$ is the curvature of the critical line $\beta_{ct}(\mu_q) \simeq \beta_{ct}(0) - c\,(\mu_q/T)^2 /2$ in the $(\beta, \mu_q/T)$ plane in the chiral limit at low densities.
From the scaling relation, we find
\begin{eqnarray}
\left. \frac{d^2 M}{d(\mu_q/T)^2} \right|_{\mu_q=0} 
= c \left. \frac{dM}{dt} \right|_{\mu_q=0}, \hspace{5mm}
\left. \frac{dM/dt}{h^{1/\delta-1/\beta\delta}} \right|_{\mu_q=0} 
= \left. \frac{df(x)}{dx} \right|_{x=t/h^{1/\beta\delta}}.
\label{eq:der2sc}
\end{eqnarray}
Then, the coefficient $c$ can be obtained by 
\begin{eqnarray}
c \;\equiv\;
- \frac{d^2 \beta_{ct}}{d(\mu_q/T)^2} \;=\; 
\left. \frac{1}{h^{1/\delta-1/\beta\delta}}
\left. \frac{d^2 \langle \bar{\psi} \psi \rangle}{d(\mu_q/T)^2} \right|_{\mu_q=0} \right/ \!
\left. \frac{df(x)}{dx} \right|_{x=t/h^{1/\beta\delta}}.
\label{eq:der2scfn}
\end{eqnarray}
The curvature of the critical temperature $T_c (\mu_B)$ in physical units can now be given as 
\begin{eqnarray}
\kappa \;\equiv\; - \frac{1}{2T_c} \frac{d^2 T_c}{d(\mu_B/T)^2} \;=\; 
- \frac{1}{18 T_c} \frac{d^2 T_c}{d(\mu_q/T)^2} \;=\;  
- \frac{c}{18} \left/ \! \left. \left( a \frac{d\beta}{da} \right) \right|_{\beta=\beta_{ct}} \right. , 
\label{eq:curvature}
\end{eqnarray}
where $\mu_B=3\mu_q$ is the baryonic chemical potential 
and $a (d\beta/da)$ is the lattice beta function.

Here, a careful treatment is required with Wilson-type quarks because the chiral symmetry is explicitly broken at finite $a$. 
Following Refs.~\cite{iwasaki,cppacs1}, 
we define $m_q$ and $\langle \bar{\psi} \psi \rangle$ by axial-vector Ward-Takahashi identities \cite{bochicchio}:
$
2m_q a = -m_{\rm PS} \left. \langle \bar{A_4}(t) \bar{P}(0) \rangle \right/
\langle \bar{P}(t) \bar{P}(0) \rangle,
$
where $P$ and $A_{\mu}$ are the pseudo-scalar and axial-vector meson operators  
and the bar means the spatial average, and 
\begin{eqnarray}
\langle \bar{\psi} \psi \rangle
\;=\;  \frac{2 m_q a}{N_s^3\, N_t} \sum_{x,x'} \langle P(x) P(x') \rangle
\;=\;  \frac{2 m_q a (2K)^2 }{N_s^3 N_t} \left\langle 
{\rm tr} \left( D^{-1} \gamma_5 D^{-1} \gamma_5 \right) \right\rangle,
\end{eqnarray}
where $D$ is the quark matrix, $K$ is the hopping parameter, and $N_s^3 \, N_t$ is the number of sites. 
They satisfy the Ward-Takahashi identity,
$
\langle \partial_{\mu} A_{\mu} (x) P(x') \rangle
-2m_q a \langle P(x) P(x') \rangle
=\delta(x- x') \langle \bar{\psi} \psi \rangle,
$
in the continuum limit.

Although a direct simulation at $\mu_q \ne 0$  is difficult due to the complex weight problem, the reweighting method is applicable at small $\mu_q$. 
Alternatively, we may directly calculate operators corresponding to derivatives of observables in terms of $\mu_q$ at $\mu_q=0$.  
We test these two methods to evaluate the derivatives of $\langle \bar{\psi} \psi \rangle$.

\paragraph{Method~1: Reweighting method}
We use the reweighting method to calculate $\langle \bar{\psi} \psi \rangle$ at $\mu_q\ne0$,
\begin{eqnarray}
(2K)^2 \left\langle {\rm tr} \left( D^{-1} \gamma_5 D^{-1} \gamma_5 \right) 
\right\rangle_{\beta, \mu_q}
&=& (2K)^2 \, \frac{1}{Z} \int {\cal D} U \ {\rm tr}(D^{-1} \gamma_5 D^{-1} \gamma_5 ) 
(\det D)^{N_{\rm f}} e^{-S_g} \nonumber \\
&=& \frac{ (2K)^2 \left\langle 
{\rm tr} \left( D^{-1} \gamma_5 D^{-1} \gamma_5 \right)(\mu_q) 
e^{N_{\rm f} [ \ln \det D(\mu_q) - \ln \det D(0)]} \right\rangle_{\beta, 0}
}{\left\langle e^{N_{\rm f} [ \ln \det D(\mu_q) - \ln \det D(0)]} 
\right\rangle_{\beta, 0}}, \ \ 
\label{eq:rewchi}
\end{eqnarray}
where $N_{\rm f}=2$.
Because we calculate the second derivative with respect to $\mu_q$,
we evaluate $\ln \det M(\mu_q)$ and 
${\rm tr} ( D^{-1} \gamma_5 D^{-1} \gamma_5)$ 
by a Taylor expansion up to $O(\mu_q^2)$ \cite{BS02}:
\begin{eqnarray}
&& N_{\rm f} \left[\ln \det D(\mu_q) - \ln \det D(0)\right] = \mu_q a \, {\cal Q}_1 
+ \frac{(\mu_q a)^2}{2} {\cal Q}_2 + O(\mu_q^3), \nonumber \\
&& (2K)^2 {\rm tr} \left( D^{-1} \gamma_5 D^{-1} \gamma_5 \right)(\mu_q) 
= (2K)^2 {\rm tr} \left( D^{-1} \gamma_5 D^{-1} \gamma_5 \right)(0) 
+ \mu_q a {\cal C}_1 
+ \frac{(\mu_q a)^2}{2} {\cal C}_2 +O(\mu_q^3). \ \ \ \ \
\label{eq:dpipi}
\end{eqnarray}
where ${\cal Q}_n$ and ${\cal C}_n$ are defined by
\begin{eqnarray}
{\cal Q}_n = N_{\rm f} \frac{\partial^n \ln \det D} 
{\partial (\mu_q a)^n}, 
\hspace{5mm}
{\cal C}_n = (2K)^2 \frac{\partial^n {\rm tr} 
\left( D^{-1} \gamma_5 D^{-1} \gamma_5 \right)} {\partial (\mu_q a)^n}. 
\label{eq:basic}
\end{eqnarray}
These quark operators can be evaluated by a random noise method. 
We then fit the data by
$
\left\langle \bar{\psi} \psi \right\rangle (\mu_q)
= x + y \, (\mu_q/T)^2,
$
to extract 
$x=\left\langle \bar{\psi} \psi \right\rangle (0) $ and 
$\displaystyle{y= \frac{1}{2}\frac{d^2 \left\langle \bar{\psi} \psi \right\rangle}{d (\mu_q/T)^2} (0)}$. 
Here, the first derivative is zero due to the symmetry $\mu_q \to -\mu_q$.

\paragraph{Method~2: Direct calculation of derivative operators}
Derivatives of $\langle \bar{\psi} \psi \rangle$ are given by
\begin{eqnarray} 
\left\langle \bar{\psi} \psi \right\rangle
\biggr|_{\mu_q =0}
&=& \frac{2m_q a}{N_s^3 N_t} {\cal F}_0 , \hspace{8mm} 
\frac{\partial \left\langle \bar{\psi} \psi \right\rangle
}{\partial (\mu_q/T)}  \biggr|_{\mu_q =0} 
= 
\frac{2m_qa}{N_s^3 N_t^2} 
\left( {\cal F}_1 - {\cal F}_0 {\cal A}_1 \right) =0 , \nonumber \\
\frac{\partial^2 \left\langle \bar{\psi} \psi \right\rangle
}{\partial (\mu_q/T)^2}  \biggr|_{\mu_q=0} 
&=& 
\frac{2m_q a}{N_s^3 N_t^3} 
\left( {\cal F}_2 -2 {\cal F}_1 {\cal A}_1 - {\cal F}_0 {\cal A}_2 
+2 {\cal F}_0 {\cal A}_1^2 \right) 
= \frac{2m_q a}{N_s^3 N_t^3} 
\left( {\cal F}_2 - {\cal F}_0 {\cal A}_2 \right) ,
\end{eqnarray}
at $\mu_q=0$, where 
\begin{eqnarray}
&& {\cal A}_1 = \left\langle {\cal Q}_1 \right\rangle, \hspace{4mm}
{\cal A}_2 = \left\langle {\cal Q}_2 \right\rangle 
+\left\langle {\cal Q}_1^2 \right\rangle, \hspace{4mm}
{\cal F}_0 = 
\left\langle {\cal C}_0 \right\rangle, \nonumber \\ &&
{\cal F}_1 = \left\langle {\cal C}_1 \right\rangle 
+ \left\langle {\cal C}_0 {\cal Q}_1 \right\rangle, \hspace{4mm}
{\cal F}_2 = \left\langle {\cal C}_2 \right\rangle 
+ 2 \left\langle {\cal C}_1 {\cal Q}_1 \right\rangle 
+ \left\langle {\cal C}_0 {\cal Q}_2 \right\rangle 
+ \left\langle {\cal C}_0 {\cal Q}_1^2 \right\rangle.
\end{eqnarray}
Note that ${\cal A}_n$ and ${\cal F}_n$ are zero for odd $n$'s at $\mu_q =0$.

\section{Numerical results}

We perform finite-temperature simulations of two-flavor QCD at  $\mu_q=0$ on a $16^3 \times 4$ lattice and combine them with configurations obtained in Refs.~\cite{whot07,whot09,whot13}. 
The RG-improved Iwasaki gauge action and the 2-flavor clover-improved Wilson quark action are adopted. The measurements are done every 10 trajectories and 500 configurations are used for the analysis at each simulation point. 
The range of $\beta$ is 1.5 to 2.0.
The quark mass $m_q a$ is computed performing zero temperature simulations 
on $16^3 \times 24$ or $16^4$ lattices at the same points as the $T \neq 0$ simulations.
The number of configurations used for the measurement is 50 for $16^4$ and 200 for $16^3 \times 24$. 
The pseudo-scalar to vector meson mass ratio at $T=0$ is  
$m_{\rm PS}/m_{\rm V} \approx 0.65$, $0.70$ and $0.80$.
$\langle \bar{\psi} \psi \rangle$ and its derivatives are calculated by the random noise method \cite{whot09,whot13} 
using 150 noise vectors for each color and spin. 

\subsection{O(4) scaling at $\mu_q=0$}
\label{sec:zeromu}

We first test the O(4) scaling relation
$
M/h^{1/\delta} = f(t/h^{1/\beta \delta})
$
assuming the O(4) critical exponents. 
The scaling function $f(x)$ for O(4) spin model is given in Ref.~\cite{engels00,engels03}. 
We adjust three fit parameters to fit the QCD data to $f(x)$, 
\textit{i.e.}, the critical point $\beta_{ct}$ at $\mu_q=0$ 
and the scales of the horizontal and vertical axes of the scaling function.
We also vary the range of data used in the fit.
The best fit result is shown in Fig.~\ref{fig2}.
The black curve is the O(4) scaling function. 
Only the data of $m_{\rm PS}/m_{\rm V} \simeq 0.65$ (red) and $0.70$ (blue) in the range $\beta \leq 1.75$ are used in the fit, 
while all data including those at $\beta > 1.75$ and at $m_{\rm PS}/m_{\rm V} \simeq 0.80$ (green) are shown in the figure.
From the fit, we obtain $\beta_{ct}=1.532(32)$ for the critical point in the chiral limit.
We find that the $m_q$- and $T$-dependences of the chiral condensate in two-flavor QCD are well consistent with the O(4) universality.

\begin{figure}[tb]
\begin{minipage}{7cm}
\centering
\vskip -0.3cm
\includegraphics[clip,width=6cm]{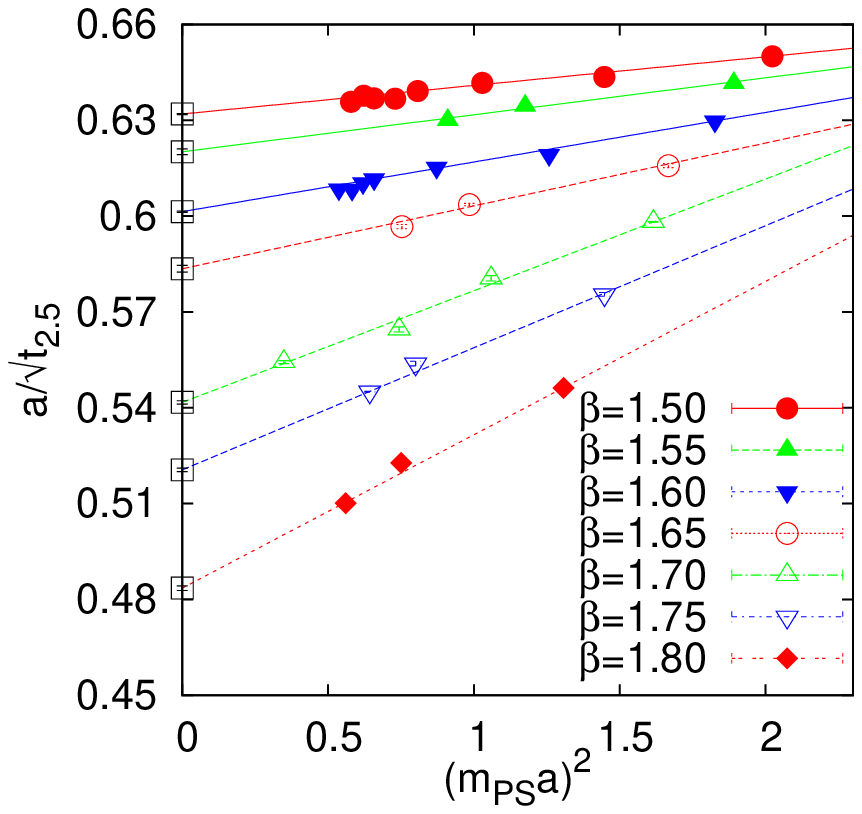}
\vskip -0.3cm
\caption{
Chiral extrapolation of $a/\!\sqrt{t_{2.5}}$.
}
\label{fig4}
\end{minipage}
\hspace{8mm}
\begin{minipage}{7cm}
\centering
\vskip -0.3cm
\includegraphics[clip,width=6cm]{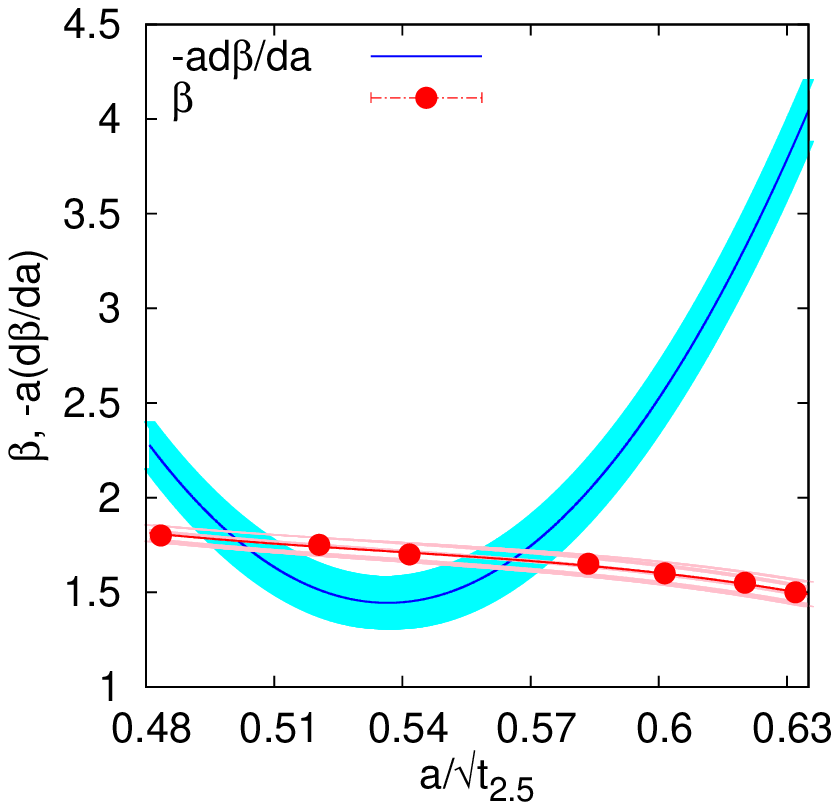}
\vskip -0.3cm
\caption{
The beta function in chiral limit.
}
\label{fig5}
\end{minipage}
\end{figure}

\subsection{Lattice scale and beta function in the chiral limit}
\label{sec:betafunc}
We determine the lattice spacing $a$ by the gradient flow method \cite{Luscher1,Luscher2}
using the $T=0$ configurations obtained on $16^4$ and $16^3 \times 24$ lattices. 
We measure $t^{2} \langle E(t) \rangle$ as function of the flow time $t$, 
where $E(t)$ is the gauge energy density defined as 
$ E(t) = \frac{1}{4} G^{a}_{\mu \nu} (t) G^{a}_{\mu \nu} (t)$ 
using the field strength $G^a_{\mu \nu}(t)$ of flowed gauge field at $t$.
To reduce the lattice discretization error in $\langle E \rangle$, 
we construct the square of the field strength by an appropriate linear combination 
of the plaquette and clover operators following Ref.~\cite{Oa2improve}:
$t^2 \langle E \rangle = (-4c_1+1/4) t^2 \langle E^{\rm pl} \rangle +(4c_1+3/4) t^2 \langle E^{\rm cl} \rangle$,
where $E^{\rm pl}$ and $E^{\rm cl}$ are defined by the plaquette and clover, respectively, and $c_1 = -0.331$ is the improvement parameter of the Iwasaki gauge action.
Because $t^{2} \langle E \rangle$ is dimension-less, it can depend on $t$ and $a$ only through their dimension-less ratio $\sqrt{t}/a$.
We determine the lattice spacing $a$ in a unit of the flow time $\sqrt{t_{X}}$ at which $t^{2} \langle E \rangle=X$.
In this study, we test five values, $X=1.5$, 2.0, 2.5, 3.0 and 3.5.

The results of $a/\!\sqrt{t_{X}}$ are plotted in Fig.~\ref{fig4} for $X=2.5$.
At each $\beta$, we extrapolate $a/\!\sqrt{t_{X}}$ to the chiral limit where the pseudo-scalar meson mass $m_{PS}$ vanishes, by fitting the data to the fitting function $a/\!\sqrt{t_{X}} = A (m_{PS}a)^2 + B$ with fit parameters $A$ and $B$.
The red symbols in Fig.~\ref{fig5} are $\beta(a/\!\sqrt{t_{2.5}})$ in the chiral limit.
We then fit $\beta$ by a cubic function of $a/\!\sqrt{t_{X}}$
to obtain $a(d\beta/da)$ in the chiral limit by differentiating the fit result $\beta(a/\!\sqrt{t_{X}})$. 
The result of $a(d\beta/da)$ for $X=2.5$ is shown by the blue curve in Fig.~\ref{fig5}.
We find that $a(d\beta/da)$ is well stable in the range $X=2.0$ - 3.0.
We summarize the results for the beta function at $\beta_{ct}$ in Table \ref{tab2}.

\begin{figure}[tb]
\begin{minipage}{7cm}
\centering
\vskip -0.3cm
\includegraphics[clip,width=6.4cm]{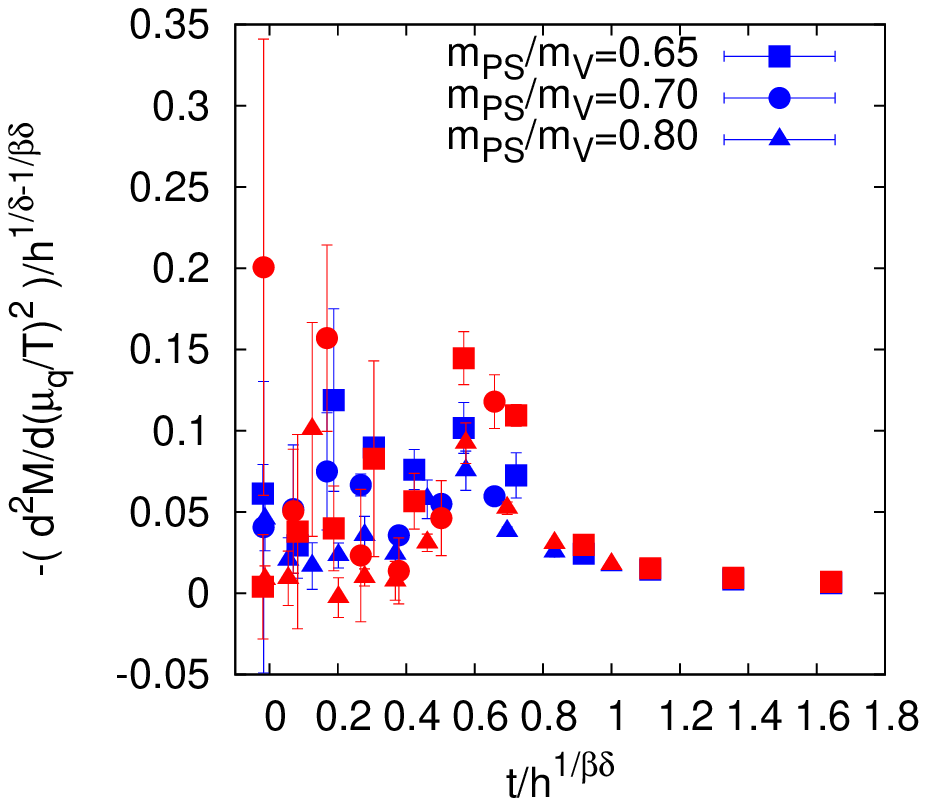}
\vskip -0.3cm
\caption{
The second derivative of the chiral condensate.
}
\label{fig3a}
\end{minipage}
\hspace{8mm}
\begin{minipage}{7cm}
\centering
\vskip -0.2cm
\includegraphics[clip,width=6cm]{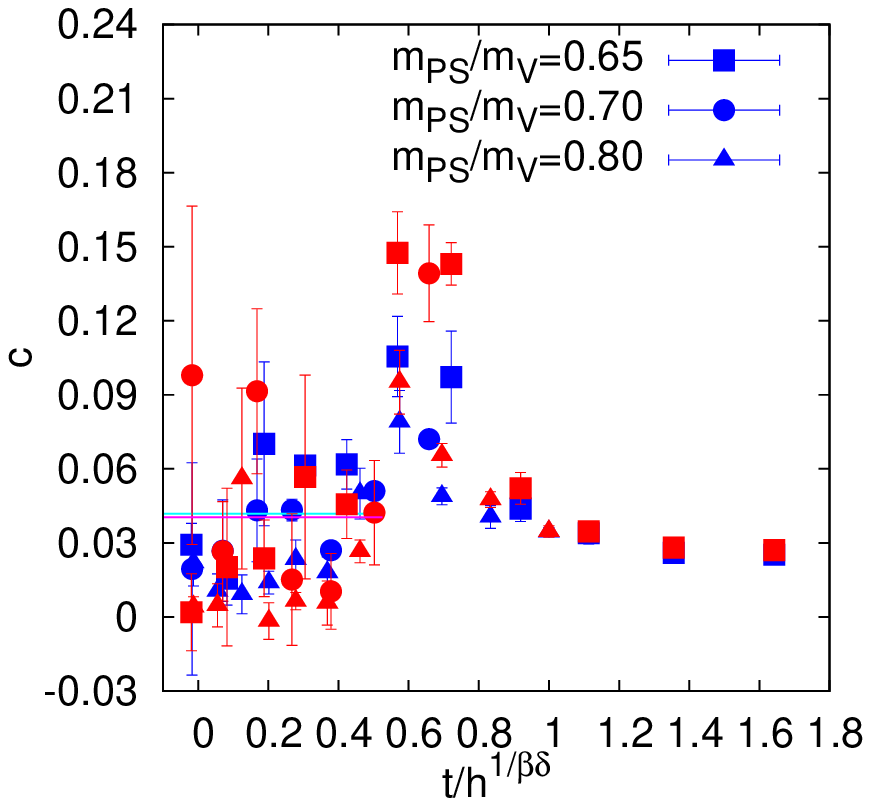}
\vskip -0.3cm
\caption{
Curvature of the chiral transition/ crossover line
in the $(\beta, \mu_q/T)$ plane.
}
\label{fig3}
\end{minipage}
\end{figure}

\begin{figure}[tb]
\begin{minipage}{14.6cm}
\centering
\makeatletter
\def\@captype{table}
\makeatother
 \footnotesize
 \begin{tabular}{|c|ccc|ccc|}
 \cline{1-7}
 \multicolumn{1}{|c|}{ } &
 \multicolumn{3}{c|}{Method~1} &
 \multicolumn{3}{c|}{Method~2} \\
 \cline{1-7}
 \cline{1-7}
 \multicolumn{1}{|c|}{$c$} &
 \multicolumn{3}{c|}{0.0418(22)} &
 \multicolumn{3}{c|}{0.0404(60)} \\
 \cline{1-7}
 \multicolumn{1}{|c|}{scale} &
 \multicolumn{1}{c}{$\sqrt{t_{2.0}}/a$} &
 \multicolumn{1}{c}{$\sqrt{t_{2.5}}/a$} &
 \multicolumn{1}{c|}{$\sqrt{t_{3.0}}/a$} & 
 \multicolumn{1}{c}{$\sqrt{t_{2.0}}/a$} &
 \multicolumn{1}{c}{$\sqrt{t_{2.5}}/a$} &
 \multicolumn{1}{c|}{$\sqrt{t_{3.0}}/a$} \\ 
 \cline{1-7}
 $-a(d\beta/da)$ & 3.90(25) & 3.85(16) & 4.01(39) & 3.90(25) & 3.85(16) & 4.01(39) \\
 $\kappa$ & 0.00060(5) & 0.00060(4) & 0.00058(6) & 0.00057(9) & 0.00058(9) & 0.00056(10) \\
 \cline{1-7} 
 \cline{1-7}
\end{tabular}
\vspace{-0.3mm}
\caption{The beta function at $\beta_{ct}=1.532(32)$  
and the curvatures $c$ and $\kappa$ of the chiral transition line in the chiral limit at $\mu_q=0$.}
\label{tab2}
\end{minipage}
\end{figure}

\subsection{Curvature of the chiral transition line in the chiral limit}
\label{sec:curvature}

We now evaluate the curvature of the chiral transition line in the chiral limit at low densities, by applying the two methods discussed in Sec.~\ref{sec:method}.
Our results for the second derivative of the chiral condensate (\ref{eq:der2sc}) are shown in Fig.~\ref{fig3a}.
O(4) exponents and the O(4) scaling function are assumed.
The shapes of the symbols, square, circle and triangle, correspond to $m_{\rm PS}/m_{V}=0.65$, $0.70$ and $0.80$, respectively.
The blue (red) symbols are the results of the Method~1 (Method~2).
We find that the results of the two method are well consistent with each other.
The curvature $c$ in the $(\beta, \mu_q/T)$ plane is given by Eq.~(\ref{eq:der2scfn}).
The result of $c$ is plotted in Fig.~\ref{fig3}. 
To evaluate $c$, we perform a constant fit using the data of $m_{\rm PS}/m_{V}=0.65$ and $0.70$ in the fit range of $\beta \leq 1.75$.
We obtain $c = 0.0418(22)$ by the Method~1, and $c = 0.0404(60)$ by the Method~2, 
as shown by blue and red lines in Fig.~\ref{fig3}.

Combining the results of $c$ and that of $a(d\beta/da)$ obtained in Sec.~\ref{sec:betafunc}, we calculate the curvature $\kappa$ of the critical temperature $T_c (\mu_q)$ in physical units by Eq.~(\ref{eq:curvature}).
Our results of $\kappa$ are summarized in Table~\ref{tab2}. 
The difference between the results of Method~1 and Method~2 turned out to be much smaller than the statistical errors. 
From these results, we obtain $\kappa =0.0006(1)$ for two-flavor QCD.

\section{Summary and discussions}
\label{sec:summary}

Performing finite-temperature simulations of two-flavor QCD with improved Wilson quarks,  
we evaluated the curvature of the chiral transition line $T_c(\mu_B)$ in the chiral limit at low densities.
We first confirmed that the chiral order parameter at zero-density is consistent with the O(4) scaling suggested by the standard scenario. 
Extending the O(4) scaling to small chemical potentials, we then calculated the second derivative of the chiral condensate in terms of the chemical potential, which is needed to evaluate the curvature, by adopting two methods -- a reweighting method and a direct calculation of derivative operators.
We found that the results of the two methods agree with each other.
Combining the results with a study of the lattice beta function by a gradient flow method, 
we obtain $\kappa =0.0006(1)$ for the curvature in two-flavor QCD.
This value is much smaller than the result obtained with improved staggered quarks in (2+1)-flavor QCD~\cite{bnl-bie10}, which is $\kappa_q  =0.059(2)(4)$, \textit{i.e.} $\kappa = \kappa_q/9 = 0.0066(2)(4)$.
To clarify if the difference is due to a strange quark effect and/or the choice of the lattice quark action, a systematic study varying quark mass as well as lattice spacing and volume is needed.

\paragraph{Acknowledgments}
We thank other members of the WHOT-QCD Collaboration for valuable discussions.
This work is in part supported by JSPS KAKENHI Grant 
(Nos.\ 25800148, 26287040, 26400244, 26400251, 15K05041).
Computations are performed at High Energy Accelerator Research Organization (KEK) 
(Nos.\ 14/15-23, 15/16-25, 15/16-T06), at Center for Computational Sciences, University of Tsukuba, 
and at Yukawa Institute, Kyoto University.

\end{document}